# Experimental non-classicality of an indivisible quantum system

Radek Lapkiewicz[1,2], Peizhe Li[1], Christoph Schaeff[1,2], Nathan K. Langford[1,2,*], Sven Ramelow[1,2], Marcin Wieśniak[1,†], and Anton Zeilinger[1,2]

[1]*Vienna Center for Quantum Science and Technology (VCQ), Faculty of Physics, University of Vienna, Boltzmanngasse 5, Vienna A-1090, Austria*
[2]*Institute for Quantum Optics and Quantum Information (IQOQI), Austrian Academy of Sciences, Boltzmanngasse 3, Vienna A-1090, Austria*

Quantum theory demands that, in contrast to classical physics, not all properties can be simultaneously well defined. The Heisenberg Uncertainty Principle is a manifestation of this fact. Another important corollary arises that there can be no joint probability distribution describing the outcomes of all possible measurements, allowing a quantum system to be classically understood. We provide the first experimental evidence that even for a single three-state system, a qutrit, no such classical model can exist that correctly describes the results of a simple set of pairwise compatible measurements. Not only is a single qutrit the simplest system in which such a contradiction is possible, but, even more importantly, the contradiction cannot result from entanglement, because such a system is indivisible, and it does not even allow the concept of entanglement between subsystems.

## 1. Introduction

The Heisenberg Uncertainty Principle is perhaps one of the most curious and surprising features of quantum physics: it prohibits certain properties (e.g. position and momentum of a single particle) from being simultaneously well defined[1]. Such incompatible properties, however, contrast strongly with what we experience in our everyday lives. If we look at a globe of the world, we can only see one hemisphere at any given time, but we suppose that the shapes of the continents on the far side remain the same irrespective of the observer's vantage point. Thus, by spinning the globe around to view different continents, we are able to construct a meaningful picture of the whole. It is reasonable to assume that observation reveals features of the continents that are present independent of which other continent we might be looking at. In this way, classical physics allows us to assign properties to a system without actually measuring it. All these properties can be assumed to exist in a consistent way, whether they are measured or not.

The world view in which system properties are defined independently of both their own measurement and what other measurements are made, we call non-contextual realism. From this viewpoint, mathematically, there must be a joint probability distribution for these properties, defining the outcome probabilities for an experiment in which they are observed (if a joint probability distribution exists, rolling appropriately weighted dice would reproduce all behavior of such an experiment). The reverse is not necessarily true. Nature could in principle be such that while joint probability distributions exist they do not relate to properties of a system.

Importantly, any non-contextual hidden variable model provides a joint probability distribution of all measurement outcomes[2,3]. By contrast, the theoretical results of, e.g., Specker[4], Bell[5], and Kochen and Specker[6], preceded by findings of Gleason[7], imply that such models are in conflict with quantum mechanics. Simpler cases where that conflict occurs have also been found[8,9,10,11,12], some of which were tested experimentally with pairs of qubits (realized by two degrees of freedom of a single particle)[13,14]. Recently, an inequality satisfied by non-contextual hidden variable models and violated by quantum mechanics for all states of two qubits was introduced[15] and tested experimentally[16,17,18].

Klyachko, Can, Binicioğlu and Shumovsky (KCBS)[19] studying the question of contextuality for individual spin-1 particles proposed an elegant inequality. This inequality actually has a much broader significance as it should apply to any five two-outcome measurements. Here, we violate this inequality and show that this thereby provides the first experimental evidence that, for a single three-state system, no joint probability distribution – and therefore no non-contextual theory – can exist[20,21]. This is the simplest system in which such a contradiction is possible. Perhaps even more interestingly, the contradiction cannot result from entanglement, because a single three-state system is indivisible and it does not even allow the concept of entanglement.

---

[*]Present address: Clarendon Laboratory, Department of Physics, University of Oxford, Parks Road, Oxford, OX13PU, UK.
[†]Present address: Institute of Theoretical Physics and Astrophysics, University of Gdansk, PL-80-952 Gdansk, Poland.

## 2. KCBS inequality

Consider five numbers $a_1$, $a_2$, $a_3$, $a_4$ and $a_5$, each equal to +1 or −1. For any choice of them, the following algebraic inequality is true[19]:

$$a_1 a_2 + a_2 a_3 + a_3 a_4 + a_4 a_5 + a_5 a_1 \geq -3 \qquad (1)$$

Let these numbers now be the results of five corresponding two-outcome measurements $A_1$, $A_2$, $A_3$, $A_4$, and $A_5$. Then, assuming there exists a joint probability distribution for the $2^5$ possible measurement outcome combinations, taking the average of (1) gives (see Supplementary Information (SI) 1 for more details):

$$\langle A_1 A_2 \rangle + \langle A_2 A_3 \rangle + \langle A_3 A_4 \rangle + \langle A_4 A_5 \rangle + \langle A_5 A_1 \rangle \geq -3 \qquad (2)$$

Here $\langle .. \rangle$ designates averages of measurement outcomes and does not necessarily imply a quantum-mechanical expectation value.

We would like to emphasize that an experimental violation of the inequality (2), excludes - without any further assumptions - the description of the measurement results using a joint probability distribution. Note that it also excludes any non-contextual realistic model for the results, because any such model would allow the construction of a joint probability distribution.

## 3. Experimental scheme

The above inequality can be tested experimentally if we assume that a long series of individual experimental runs would result in a fair sampling of a joint probability distribution (if one were to exist). Our experimental implementation consists of five main stages, depicted in Fig. 1.b-1.f. We prepare single photons, each distributed among three modes (see Fig. 1.a.) monitored by detectors. At each stage the two outcomes of a given measurement are defined by asking, *"did the corresponding detector click?"* For example, at stage 1 (see Fig. 1.b), the outcomes of the measurement $A_1$ are given by the response of the upper detector, and we assign numbers to the outcomes as above: i.e. $A_1 = -1$ if it clicked, and $A_1 = +1$ if not. Similarly, the outcomes of measurement $A_2$ are given by the response of the lower detector. By measuring $A_1$ and $A_2$ together for a number of photons we obtain the average value $\langle A_1 A_2 \rangle$, the first term of the inequality (2).

To move to the second stage, we perform a transformation ($T_1$) on the two upper modes (see Fig. 1.c). Because the mode monitored by the lower detector is not affected by $T_1$, this detector still measures the outcome of $A_2$. The upper detector, however, defines a different measurement – we call it $A_3$. Most significantly, for any specific run of the experiment it appears reasonable to assume that whether or not detector $A_2$ clicks must be independent of whether we apply the transformation to the other two modes or not. In the remaining three stages, we apply three more transformations, each time changing one measurement and leaving the other unaffected.

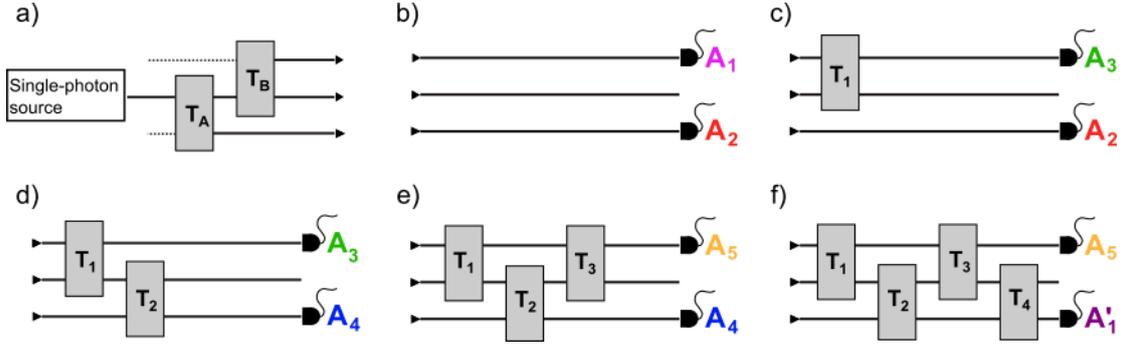

**Figure 1 | The conceptual scheme of the experiment with the preparation and five successive measurement stages.** Straight, black lines represent the optical modes (beams), gray boxes represent transformations on the optical modes. **a)** Single photons are distributed among three modes by transformations $T_A$ and $T_B$. This preparation stage is followed by one of the five measurement stages. **b)-f)** At each stage, the response of two detectors monitoring the optical modes defines a pair of measurements as appearing in inequality (2). Outcomes of the measurements are defined by asking, "did the corresponding detector click?" If a detector clicks (does not click), then a value of -1 (+1) is assigned to the corresponding measurement. A key aspect of our experimental implementation is that each transformation acts only on two modes, leaving the other mode completely unaffected. Thus, the part of the physical setup corresponding to the measurement of $A_2$ is exactly the same in Fig. 1.b and Fig. 1.c. Note that this setup can also be arranged so that the choice between $A_1$ and $A_3$ is made long after $A_2$ is measured. Thus it appears reasonable to assume that the measurement result for $A_2$ is independent of whether it is measured together with $A_1$ or $A_3$. The same reasoning can be applied to the measurements $A_3$, $A_4$, and $A_5$.

The last transformation ($T_4$) is chosen such that the new measurement should be equivalent to the original measurement of $A_1$. Unlike for the other measurements, however, this new measurement apparatus (Fig. 1.f) is not physically the same as the one measuring $A_1$ in the Fig. 1.b. We therefore call the sixth measurement $A'_1$ and derive the following new inequality to replace inequality (2) (see Supplementary Information for details),

$$\langle A_1 A_2 \rangle + \langle A_2 A_3 \rangle + \langle A_3 A_4 \rangle + \langle A_4 A_5 \rangle + \langle A_5 A'_1 \rangle \geq -3 - \varepsilon \qquad (3)$$

with $\varepsilon = 1 - \langle A'_1 A_1 \rangle$ - note that for the ideal case of $A'_1 = A_1$ this extended inequality (3) would reduce to the original one (2). The measurements for $A_1$ and $A'_1$ occur at different stages of the experiment, so the expectation value $\langle A'_1 A_1 \rangle$ cannot be measured in the same way as the other terms. Instead it is measured by realizing that in the ideal case, whenever detector $A_1$ fires $A'_1$ must also fire and vice versa. Therefore, if the upper beam is blocked where $A_1$ would be measured, then $A'_1$ should never click. Likewise, blocking the other two modes should not change the count rate at $A'_1$. Therefore we can rewrite $\langle A'_1 A_1 \rangle$:

$$\langle A'_1 A_1 \rangle = 1 - 2(P(A'_1 = -1 | A_1 = 1) P(A_1 = 1) + P(A'_1 = 1 | A_1 = -1) P(A_1 = -1)) \qquad (4)$$

with the respective conditional probabilities experimentally accessible by blocking and monitoring the appropriate modes (see Fig. 2).

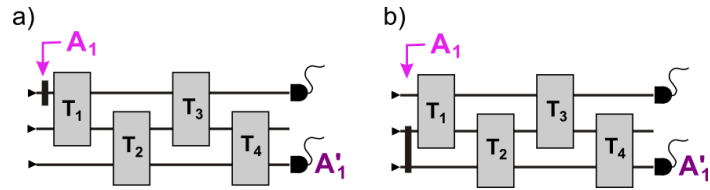

**Figure 2 | Conceptual scheme of the additional blocking stages for the measurement of $\varepsilon = 1 - \langle A'_1 A_1 \rangle$.** As in Fig. 1, straight, black lines represent optical modes (beams). Gray boxes represent transformations. Vertical black lines represent mode blocks, which absorb the photons in the mode(s) they are inserted into. If, with a block placed before the transformation

$T_1$, a photon still reaches one of the detectors, we conclude that it did not hit the block (i.e., a detector positioned in place of the block would not have clicked). Thus, **a)** having a block in the uppermost mode allows us to measure the conditional probability $P(A'_1 = -1 | A_1 = 1)$, while **b)** blocking the two remaining modes allows us to measure $P(A'_1 = 1 | A_1 = -1)$.

In our measurements, we find that $\varepsilon = 0.081(2)$, thus bounding the left-hand side of inequality (3) by $-3.081(2)$, and that each of the terms on the left-hand side is less than $-0.7$, adding to give $-3.893(6)$. This represents a violation of inequality (3) by more than 120 standard deviations, demonstrating that no joint probability distribution is capable of describing our results. The violation also excludes any non-contextual hidden variable model. The result does, however, agree well with quantum-mechanical predictions, as we will show now.

### 4. Quantum-mechanical description

A single photon distributed among three modes can be described by the mathematical formalism used for spin-1 particles. Using this formalism, the measurements performed in our experiment can be expressed by spin operators as follows $A_i = 2\hat{S}_i^2 - 1$. $\hat{S}_i$ is a spin projection onto the direction $\vec{l}_i$ in real three-dimensional space. Two measurements $\hat{S}_i^2$ and $\hat{S}_j^2$ are compatible if and only if the directions $\vec{l}_i$ and $\vec{l}_j$ are orthogonal. Thus, the five measurement directions have to be pairwise orthogonal to make the measurements themselves pair-wise compatible (Fig. 3). In our experiment we have three modes which by design represent orthogonal states. These can be seen as orthogonal directions in the spin case. An essential feature of a spin-1 system is that out of three projections squared onto three orthogonal directions exactly two are equal to 1 and the remaining one must be equal to 0. Correspondingly, a single photon distributed among three different modes will cause a click in exactly one detector only and none in the other two.

For the maximal violation of inequalities (2) and (3), the five directions form a regular pentagram and the input state has zero spin along the symmetry axis of the pentagram[19]. In the ideal case, i.e., when $A_1 = A'_1$ ($\varepsilon = 0$), if the optimal state is taken, quantum mechanics predicts the value of the left-hand side of the inequality (3) to be $5 - 4\sqrt{5} \approx -3.944$. Smaller violations are also predicted for a range of non-ideal pentagrams and other input states. In our case, the departure from the maximum achievable violation can be attributed to residual errors in the settings of experimental parameters (see Methods).

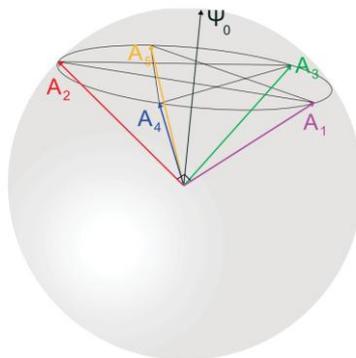

**Figure 3 | Representation of the measurements and a state providing maximal violation of the inequality (2) by directions in three-dimensional space[19].** The measurement directions are labeled by the measurements $A_i$ (i=1,2...,5). They are given as $2\hat{S}_i^2 - 1$, where $\hat{S}_i$ is a spin projection onto the direction $\vec{l}_i$. The five measurement directions are pair-wise orthogonal, making the measurements $A_i$ pair-wise compatible. These five pairs correspond to the five measurement devices from Figs 1.b-1.f. For a maximal violation, the directions form a regular pentagram and the input state $\Psi_0$ has zero spin along the symmetry axis of the pentagram.

The five pairs of compatible measurements correspond to the five measurement devices described in Figs 1.b-1.f. Spin rotations, necessary to switch between various measurement bases, can be realized by combining the optical modes e.g., on a tunable beam splitter. Each measurement basis rotation leaves one of the axes unaffected, because only two modes are combined and one is untouched. The measurements are pair-wise compatible, because the overlap of the modes is negligible and measurements performed on one mode therefore do not affect those performed on the other.

## 5. Experimental realization of the scheme

We realize the above scheme using a single photon propagating in three modes. Two of the three modes are realized as two polarizations of a single spatial mode. Conveniently, two-mode transformations can then be implemented using half-wave plates acting on the two polarization modes propagating in the same spatial mode. Different spatial modes are combined using calcite crystals (acting as polarizing beam splitters). Thus we are able to apply transformations to any pair of modes (see Fig. 4).

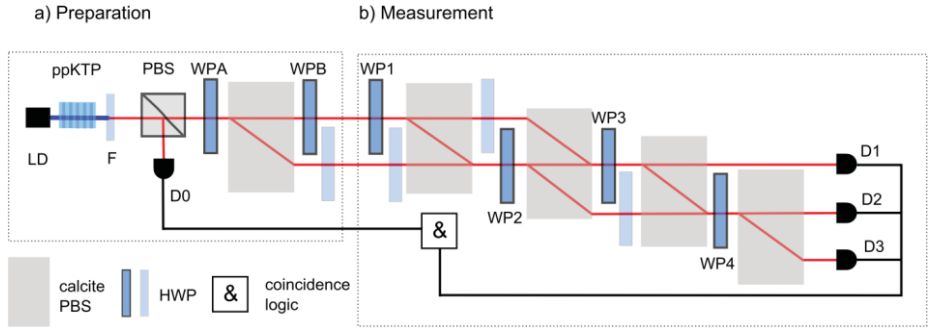

**Figure 4 | Experimental setup. a) Preparation of the required single-photon state.** A grating-stabilized laser diode at 405 nm (LD) is used to pump the nonlinear crystal (20 mm long, periodically poled Potassium Titanyl Phosphate - ppKTP) producing pairs of orthogonally polarized photons via spontaneous parametric down-conversion. The pump is filtered out with the help of a combination of dichroic mirrors and interference filters (labeled jointly as F). The photon pairs are split up at a polarizing beam splitter (PBS). Detecting the reflected photon heralds the transmitted one. Half-wave plates $WP_A$ and $WP_B$ transform the transmitted photon into the desired three-mode state. Calcite polarizing beam splitters separate and combine orthogonally polarized modes. In the measurement apparatus **b)** half-wave plates $WP_1$-$WP_4$ realize the transformations $T_1$-$T_4$ on pairs of modes (wave-plate orientations are listed in Supplementary Table 1 in SI 4). Each transformation can be "turned off" by setting the optical axis of the corresponding wave plate vertically (at *0º*). The unlabeled (light blue) wave plates serve to balance the path lengths and to switch between horizontal and vertical polarization (the second unlabeled wave plate is set to *0º*, the rest is set to *45º*). Detecting heralded single photons means in practice registering coincidences between single photon detectors: $D_0$ and each of $D_1$, $D_2$, and $D_3$. Registrations in two of the detectors $D_1$, $D_2$, and $D_3$ give the values $A_i$ necessary for the inequality (see Table 1). The third detector is used to identify the trials when the photon is lost. Note that the assignment of measurements to detectors in the experimental setup differs in some cases from that described in the simplified conceptual scheme (Fig. 1).

The experiment consists in total of seven stages. The first five stages (corresponding to the five main terms of the inequality (3)) are the measurement configurations illustrated in Fig. 1.b-1.f, while the last two depicted at Fig. 2 give us the value of $\varepsilon$. All of these measurements are realized with a single experimental apparatus tuned to one of seven configurations. The configurations 1-5 differ in the number of transformations that are "active". We activate and deactivate the transformations by changing the orientation of wave plates (see Supplementary Table 1 in SI 3 for the specific settings). For the two measurements in the final stage, where we measure conditional probabilities by blocking the appropriate modes (Fig. 2), we insert a polarizer in two orthogonal orientations between wave plates $WP_B$ and $WP_1$.

For each measurement, we recorded clicks for one second, registering about 3,500 heralded single photons. We repeated each stage 20 times, averaged the results and calculated the standard deviation of the mean to estimate the standard uncertainties which we then propagated to the final results (presented in Table 1). Due to photon loss, sometimes no photon is detected in the measurement, despite the observation of a heralding event. We therefore discard all events for which only the trigger detector ($D_0$) and none of the measurement detectors ($D_1$, $D_2$, $D_3$) fired. We assume that the photons we do detect are a representative sample of all created photons (fair-sampling assumption).

**Table 1 | Collected experimental results.**

a)

| D₁ | | D₂ | | D₃ | | Calculated contribution | |
|---|---|---|---|---|---|---|---|
| condition | value | condition | value | condition | value | term | value |
| $P(A_1=1, A_2=-1)$ | 0.471(3) | $P(A_1=-1, A_2=1)$ | 0.432(3) | $P(A_1=1, A_2=1)$ | 0.097(1) | $\langle A_1 A_2 \rangle$ | -0.805(2) |
| $P(A_2=-1, A_3=1)$ | 0.473(4) | $P(A_2=1, A_3=1)$ | 0.098(2) | $P(A_2=1, A_3=-1)$ | 0.429(4) | $\langle A_2 A_3 \rangle$ | -0.804(3) |
| $P(A_3=1, A_4=1)$ | 0.146(2) | $P(A_3=1, A_4=-1)$ | 0.429(2) | $P(A_3=-1, A_4=1)$ | 0.426(2) | $\langle A_3 A_4 \rangle$ | -0.709(3) |
| $P(A_4=1, A_5=-1)$ | 0.466(2) | $P(A_4=-1, A_5=1)$ | 0.439(2) | $P(A_4=1, A_5=1)$ | 0.095(1) | $\langle A_4 A_5 \rangle$ | -0.810(2) |
| $P(A_5=-1, A'_1=1)$ | 0.469(2) | $P(A_5=1, A'_1=-1)$ | 0.414(2) | $P(A_5=1, A'_1=1)$ | 0.117(2) | $\langle A_5 A'_1 \rangle$ | -0.766(3) |
| | | | | | | Σ | -3.893(6) |

b)

| D₁ | D₃ | D₁ + D₃ | | D₂ | | Calculated contribution | |
|---|---|---|---|---|---|---|---|
| | | condition | value | condition | value | term | value |
| 0.788(2) | 0.196(2) | $P(A'_1=1\|A_1=1)$ | 0.983(1) | $P(A'_1=-1\|A_1=1)$ | 0.017(1) | $-3-\varepsilon$ | -3.081(2) |
| 0.010(1) | 0.062(2) | $P(A'_1=1\|A_1=-1)$ | 0.072(2) | $P(A'_1=-1\|A_1=-1)$ | 0.928(2) | | |

The value columns contain the measured probabilities corrected for relative efficiencies (see Methods). Estimates of standard uncertainties (standard deviations of the means) are given in the brackets. The condition columns contain assigned measurement values (in label brackets) corresponding to heralded single-click events. Because of low detection efficiency, we need to use a third detector. It enables us to identify and discard trials in which a photon was lost (heralded no-click events). The rates of heralded double clicks (simultaneous responses of heralding detector and two other detectors) are negligible – typically two orders of magnitude smaller than the standard deviation in the rate of heralded single clicks.

**a) Results for the KCBS inequality.** Rows 1-5 correspond to terms 1-5 of inequality (3), and are measured with the corresponding devices illustrated in Figs 1.b-1.f. The last column is given by $\langle A_i A_j \rangle = P(A_i = A_j = +1) - P(A_i = -1, A_j = +1) - P(A_i = +1, A_j = -1)$

**b) Extended bound.** Rows 1 and 2 (corresponding to Figs 2.a and 2.b) display the contributions to the conditional probabilities that are necessary to evaluate the additional terms in the extended inequality (3).

A key aspect of Kochen-Specker experiments is that the co-measured observables must commute, otherwise potentially opening up the so-called compatibility loophole[22]. The construction of measurements in our experiment enforces their compatibility (the overlap of different modes is negligible) and thus makes it immune to the compatibility loophole. Detector efficiencies and losses in the setup prevent us from closing the detection loophole. Instead we assume that the statistics of unregistered events would have been the same as observed ones (fair sampling).

## 6. Conclusions

Our experimental results are in conflict with any description of nature relying on a joint probability distribution of outcomes of a simple set of measurements. This automatically also excludes any description in terms of non-contextual hidden variable models. To our knowledge this is the first observation of such a conflict for a single three-state system, which apart from being the most basic one where such a contradiction is possible, cannot even in principle contain entanglement. For such a system, the inequality (2) involves the smallest number of measurements possible. Our result sheds new light on the conflict between quantum and classical physics.

Finally, we want to point out that any model based on a joint probability distribution can in principle be non-deterministic. The experimental exclusion of such models highlights the fact that even for a single, indivisible quantum system, allowing randomness is not sufficient to enable its description with a conceptually classical model.

## Methods

**Single-photon source.** We produced heralded single photons at 810 nm using collinear spontaneous parametric down conversion (SPDC) in a nonlinear crystal (periodically poled KTP, 20 mm long). We used ~3 mW of power from a grating-stabilized laser diode at 405 nm to pump the crystal. In the type-II SPDC process, pairs of orthogonally polarized photons were produced in a polarization product state. We separate them at a polarizing beam splitter, with detection of a vertically polarized photon heralding the horizontally polarized photon in the measurement setup. In practice, therefore, it means that we record pair-wise coincidences between the heralding detector and any one of the three detectors in the measurement apparatus (heralded single clicks). We use home-built avalanche photodiode single-photon detectors and coincidence logic. The effective coincidence window (including the jitter of the detector) is about 2.3 ns.

**Calculation of probabilities.** Because of low detection efficiency, we need to use three detectors in the measurement apparatus. This enables us to identify and discard the trials in which the heralded photon was lost (it was not detected by any of the three detectors). We assume that the lost photons would have behaved the same as the registered ones (fair sampling). The count rates are corrected for differences in detector efficiencies and losses before the detectors. We obtain the correction factors, which we call single-output relative efficiencies, from the same series of measurements that we use to measure the values of the left-hand side of the inequality (3). Our method is based on the fact that ideally the sum of the count rates at the three detectors does not depend on the settings of the apparatus (e.g. wave-plate orientations), if the three relative efficiencies are properly accounted for. In such case, only the distribution of counts among the detectors can change. Using this condition we can estimate these single-output efficiencies and calculate the corrected probabilities of detection according to

$$P(D_i) = \frac{\tilde{N}_i}{\sum_{j=1}^{3} \tilde{N}_j} \quad , \text{ where } \tilde{N}_i = \frac{N_i}{\eta_i}$$ is the measured number of heralded clicks at detector $i$ divided by the corresponding relative efficiency $\eta_i$. In our experiment, the main contribution to the differences in relative efficiencies comes from differing efficiencies of the detectors themselves.

When we register a click at a single measurement detector (a heralded click), we assign a value of -1 to the corresponding measurement and +1 to the remaining two (see Fig. 5.a). We can therefore calculate the averages $\langle A_i A_j \rangle = P(A_i = A_j = +1) - P(A_i = -1, A_j = +1) - P(A_i = +1, A_j = -1)$. Because the experiment works with individual heralded photons it is impossible for two measurement detectors to fire simultaneously (i.e., $P(A_i = A_j = -1)$). Nevertheless, in the real world, "accidental" heralded double clicks very occasionally do occur, in our experiment, their contribution was negligible (typically two orders of magnitude smaller than the standard deviation in the rate of heralded single clicks).

**Inaccuracies in the experimental settings.** In the first step, wave plates $WP_1$-$WP_4$ are set to 0°. In the second step, only the orientation of $WP_1$ is changed and only this wave plate performs a non-trivial transformation. Ideally, the subsequent wave plates should perform no transformation since their orientation is set to 0°. In reality, this is not the case since their orientation can never be set perfectly. Similar problems occur between stages two and three, and three and four. This is a technical issue rather than a fundamental one, but could in principle be treated similarly to the problem of non-identical $A_1$ and $A'_1$ measurements. When the wave plates are not set to 0°, the inaccuracies in orientation are not critical, only affecting the shape of the pentagram and thereby lowering the maximum measurable violation of the inequality. The phase between the two spatial modes should ideally be set to 0. We set the phase with the help of interference signal of a diode laser. Here, the inaccuracy does not affect our argument. It only decreases the violation that can be observed.

**Acknowledgements**

This work was supported by the ERC (Advanced Grant QIT4QAD), the Austrian Science Fund (Grant F4007), the EC (Marie Curie Research Training Network EMALI), the Vienna Doctoral Program on Complex Quantum Systems (CoQuS) and the John Templeton Foundation. We acknowledge A. A. Klyachko for detailed discussion about the KCBS proposal, M. Hentschel, M. Kacprowicz, G. J. Pryde for discussions about technical issues, A. Cabello, S. Osnaghi, H. M. Wiseman, and M. Żukowski with whom we discussed the conceptual issues, and M. Nespoli for the help during the early stages of the experiment.

**References**

1. Heisenberg, W. *The Physical Principles of Quantum Theory*, University of Chicago Press, Chicago, Ill. (1930), reprinted by Dover, New York (1950).
2. For the case of Bell inequalities, it was proven by Fine[3] that the existence of a joint probability distribution is a necessary and sufficient condition for certain hidden variable models – factorizable and stochastic or deterministic – to be constructed.
3. Fine, A. Hidden variables, joint probability, and the Bell inequalities., *Phys. Rev. Lett.* **48,** 291-295 (1982).
4. Specker, E. Die Logik nicht gleichzeitig entscheidbarer Aussagen. *Dialectica* **14**, 239-246 (1960).
5. Bell, J. S. On the problem of hidden variables in quantum mechanics. *Rev. Mod. Phys.* **38**, 447-452 (1966).
6. Kochen, S. & Specker, E. P. The problem of hidden variables in quantum mechanics. *J. Math. Mech.* **17**, 59–87 (1967).
7. Gleason, A. M. Measures on the closed subspaces of a Hilbert space. *J. Math. Mech.* **6,** 885–893 (1957).
8. Mermin, N. D. Simple unified form for the major no-hidden-variables theorems. *Phys. Rev. Lett.* 65, 3373–3376 (1990).
9. Peres, A. Two simple proofs of the Kochen-Specker theorem. *Journal of Physics A* **24,** L175-L178 (1991).
10. Peres, A. *Quantum Theory: Concepts and Methods,* Ch. 7, Kluwer, Dordrecht (1993).
11. Clifton, R. Getting contextual and nonlocal elements-of-reality the easy way. *American Journal of Physics* **61,** 443-447 (1993).
12. Cabello, A., Estebaranz, J. M. & García-Alcaine, G. Bell-Kochen-Specker theorem: A proof with 18 vectors. *Phys. Lett. A* **212,** 183-187 (1996).
13. Huang Y.-F., Li C.-F., Zhang Y.-S., Pan J.-W. & Guo G.-C. Experimental Test of the Kochen-Specker Theorem with Single Photons. *Phys. Rev. Lett.* **90**, 250401 (2003).
14. Bartosik, H., et al. Experimental test of quantum contextuality in neutron interferometry. *Phys. Rev. Lett.* **103**, 040403 (2009).
15. Cabello, A. Experimentally testable state-independent quantum contextuality. *Phys. Rev. Lett.* **101**, 210401 (2008).
16. Kirchmair, G., et al. State-independent experimental test of quantum contextuality. *Nature* **460**, 494-497 (2009).
17. Amselem, E., Rådmark, M., Bourennane, M. & Cabello, A**.** State-Independent Quantum Contextuality with Single Photons. *Phys. Rev. Lett.* **103**, 160405 (2009).
18. Moussa, O., Ryan C. A., Cory, D. G. & Laflamme, R. Testing Contextuality on Quantum Ensembles with One Clean Qubit. *Phys. Rev. Lett.* **104,** 160501 (2010)
19. Klyachko, A. A., Can, M. A., Binicioğlu, S. & Shumovsky, A. S. Simple Test for Hidden Variables in Spin-1 Systems. *Phys. Rev. Lett.* **101**, 20403 (2008).
20. Lapkiewicz, R. *et al. Most basic experimental falsification of Non-contextuality* (Poster, 13[th] Workshop on Quantum Information Processing, 2010).
21. Lapkiewicz, R. *et al. Experimental Non-Classicality of an Indivisible System*. Abstr. Q29.00008 (41st Annu. Meeting Div. Atom. Mol. Opt. Phys., American Physical Society, 2010).
22. Gühne, O., et al. Compatibility and noncontextuality for sequential measurements. *Phys. Rev. A* **81**, 22121 (2010).

# Supplementary Information
## SI 1
### Derivation of the inequality (2)

Consider the function, $f(a_1, a_2, a_3, a_4, a_5) = a_1 a_2 + a_2 a_3 + a_3 a_4 + a_4 a_5 + a_5 a_1$, where the $a_j = \pm 1$ ($j=1,2,...,5$). To find its minimum value, we wish to choose the $a_j$ to make as many terms as possible equal to -1. We can succeed for the first four terms by choosing alternating values of +1 and -1, but then, whatever value we choose for $a_1$ to start the sequence, we are left with $a_1 = a_5$, thus making the fifth term equal +1. Similarly for any other sequence, there is at least one term equal +1. Consequently, $f(a_1, a_2, a_3, a_4, a_5) \geq -3$ for all choices of $a_j$, as given in inequality (1).

Next, if the results $a_j$ of measurements $A_j$ have a joint probability distribution, then the average of $f(A_1, A_2, A_3, A_4, A_5)$ is defined by

$$\langle f(A_1, A_2, A_3, A_4, A_5) \rangle = \sum_{a_1, a_2, a_3, a_4, a_5} p(a_1, a_2, a_3, a_4, a_5) f(a_1, a_2, a_3, a_4, a_5),$$

where $p(a_1, a_2, a_3, a_4, a_5) \geq 0 \;\; \forall a_j$ and $\sum p(a_1, a_2, a_3, a_4, a_5) = 1$. Because this is a convex combination, the average of $f$ has the same lower bound of -3, and because of the linearity of the average, average of a sum is equal to the sum of individual averages. This gives rise to inequality[19] (2)

$$\langle A_1 A_2 \rangle + \langle A_2 A_3 \rangle + \langle A_3 A_4 \rangle + \langle A_4 A_5 \rangle + \langle A_5 A_1 \rangle \geq -3.$$

## SI 2
### Derivation of the inequality (3)

The derivation of inequality (3) is analogous to that of inequality (2) (see SI 1), with the function $f(a_1, a_2, a_3, a_4, a_5)$ replaced by $g(a_1, a_2, a_3, a_4, a_5, a_{1'}) = a_1 a_2 + a_2 a_3 + a_3 a_4 + a_4 a_5 + a_5 a_1 - a_1 a_{1'}$.

Although, mathematically, the new inequality is very similar to the inequality (2), physically, the last term is different from the others. Because the measurements for $A_1$ and $A'_1$ occur at different stages of the experiment, the last term cannot be measured in the same way as the other ones. We must therefore be able to write the additional term $\langle A'_1 A_1 \rangle$ in terms of quantities that are experimentally accessible. Whenever $A_1 = A'_1$ the sixth term is equal to +1, otherwise it equals -1, giving

$$\langle A'_1 A_1 \rangle = P(A_1 = A'_1) - P(A_1 \neq A'_1). \tag{SI 2.1}$$

We can expand these two probabilities using conditional ones

$$P(A_1 = A'_1) = P(A'_1 = 1 | A_1 = 1) P(A_1 = 1) + P(A'_1 = -1 | A_1 = -1) P(A_1 = -1) \tag{SI 2.2}$$

$$P(A_1 \neq A'_1) = P(A'_1 = -1 | A_1 = 1) P(A_1 = 1) + P(A'_1 = 1 | A_1 = -1) P(A_1 = -1) \tag{SI 2.3}$$

Inserting (SI 2.2) and (SI 2.3) into (SI 2.1) and using the normalization condition, we arrive at

$$\langle A'_1 A_1 \rangle = 1 - 2(P(A'_1 = -1 | A_1 = 1) P(A_1 = 1) + P(A'_1 = 1 | A_1 = -1) P(A_1 = -1))$$

The four conditional probabilities $P(A'_1 = \pm 1 | A_1 = \pm 1)$ can be accessed by blocking the appropriate modes. For instance, $P(A'_1 = 1 | A_1 = -1)$ can be measured by blocking the modes that would give -1 at the stage of $A_1$ and registering the probability for $A'_1$ to be assigned $+1$ (see Fig. 2a). We obtain an experimentally testable inequality

$$\langle A_1 A_2 \rangle + \langle A_2 A_3 \rangle + \langle A_3 A_4 \rangle + \langle A_4 A_5 \rangle + \langle A_5 A'_1 \rangle - \langle A'_1 A_1 \rangle \geq -4$$

which can be rewritten as

$$\langle A_1 A_2 \rangle + \langle A_2 A_3 \rangle + \langle A_3 A_4 \rangle + \langle A_4 A_5 \rangle + \langle A_5 A'_1 \rangle \geq -3 - \varepsilon$$

where $\varepsilon = 1 - \langle A'_1 A_1 \rangle$.

## SI 3

### Nominal orientations of the wave plates

**Supplementary Table 1 | Wave plate (WP) and polarizer orientations at all the experiment stages.**

| Measured quantity | $WP_A$/° | $WP_B$/° | $WP_1$/° | $WP_2$/° | $WP_3$/° | $WP_4$/° | Polarizer |
|---|---|---|---|---|---|---|---|
| $\langle A_1 A_2 \rangle$ | -24.0 | -58.0 | 0.0 | 0.0 | 0.0 | 0.0 | - |
| $\langle A_2 A_3 \rangle$ | -24.0 | -58.0 | 109.1 | 0.0 | 0.0 | 0.0 | - |
| $\langle A_3 A_4 \rangle$ | -24.0 | -58.0 | 109.1 | 109.1 | 0.0 | 0.0 | - |
| $\langle A_4 A_5 \rangle$ | -24.0 | -58.0 | 109.1 | 109.1 | 109.1 | 0.0 | - |
| $\langle A_5 A_{1'} \rangle$ | -24.0 | -58.0 | 109.1 | 109.1 | 109.1 | -64.1 | - |
| $P(A_{1'} = -1 \mid A_1 = 1)$ | -24.0 | -58.0 | 109.1 | 109.1 | 109.1 | -64.1 | 0.0 |
| $P(A_{1'} = 1 \mid A_1 = -1)$ | -24.0 | -58.0 | 109.1 | 109.1 | 109.1 | -64.1 | 90.0 |

Let us now show that with the settings listed in the Supplementary Table 1, our setup indeed realizes the geometry shown at the Fig. 3. If we choose our coordinate system such that the state produced by the source corresponds to the 0-eigenstate of the spin-1 component in the z direction, the directions corresponding to the measurements from the inequality (2) can be written as

$$\vec{l}_i = R^{i-1} \begin{pmatrix} \sqrt{\dfrac{5-\sqrt{5}}{10}} \\ \sqrt{\dfrac{5-\sqrt{5}}{10}} \\ \dfrac{1}{\sqrt[4]{5}} \end{pmatrix},$$

where i=1,2,…,5 and R represents a rotation around the z axis by 144°, i.e.,

$$R = \begin{pmatrix} \dfrac{1}{4}(-1-\sqrt{5}) & \sqrt{\dfrac{5-\sqrt{5}}{8}} & 0 \\ -\sqrt{\dfrac{5-\sqrt{5}}{8}} & \dfrac{1}{4}(-1-\sqrt{5}) & 0 \\ 0 & 0 & 1 \end{pmatrix}.$$

Expressed in the coordinate system $(\vec{l}_1, \vec{l}_2, \vec{l}_1 \times \vec{l}_2)$, the input state takes the form $\psi_0 = \left(\frac{1}{\sqrt[4]{5}}, \frac{1}{\sqrt[4]{5}}, \sqrt{1-\frac{2}{\sqrt{5}}}\right)$.

To realize these amplitudes we set wave plates WP$_A$ and WP$_B$ to -24° and -58°, respectively (row 1, Supplementary Table 1). The unlabeled half-wave plate below WP$_B$ compensates for the π phase-shift introduced in the other two modes by WP$_B$. Apart from polarization control and balancing the dispersion, this is the role of all the unlabeled half-wave plates. At the following stages of the experiment the original coordinate system $(\vec{l}_1, \vec{l}_2, \vec{l}_1 \times \vec{l}_2)$, represented by the unit matrix is transformed to 
$\begin{pmatrix} -\alpha & \beta & 0 \\ \beta & \alpha & 0 \\ 0 & 0 & -1 \end{pmatrix}$, 
$\begin{pmatrix} -\alpha & \beta & 0 \\ \sqrt{1-2\beta^2} & -\beta & \beta \\ -\sqrt{2\beta^2+\alpha^2} & \sqrt{1-2\beta^2} & \alpha \end{pmatrix}$, 
and 
$\begin{pmatrix} \sqrt{2\beta^2-\alpha^2} & \alpha & \sqrt{1-2\beta^2} \\ \sqrt{1-2\beta^2} & -\beta & \beta \\ \alpha & 0 & -\beta \end{pmatrix}$, 
when we rotate WP$_1$, WP$_2$, and WP$_3$, respectively, by $-\frac{\pi}{4} + \arccos\left(-\frac{1}{2}\sqrt{1+\frac{1}{\sqrt{5}}}\right) \approx 109^o$ (rows 2, 3, 4, Supplementary Table 1), Here, $\alpha = \sqrt{\frac{\sqrt{5}-1}{2}}$ and $\beta = \sqrt{\frac{3-\sqrt{5}}{2}}$. Finally, we rotate WP$_4$ by $-\frac{\pi}{2} + \arccos\left(\frac{1}{2}\sqrt{1+\frac{1}{\sqrt{5}}}\right) \approx -64^o$ and transform the original coordinate system to $\begin{pmatrix} -\beta & 0 & -\alpha \\ 0 & 1 & 0 \\ \alpha & 0 & -\beta \end{pmatrix}$. We keep those wave plate settings also for the additional blocking stages (rows 6 and 7, Supplementary Table 1).

Notice that in every matrix one row is the same (up to an irrelevant global sign) as in the previous one. It is also easily verifiable that other rows from a subsequent pair of matrices have pairwise overlaps of $\pm\frac{1}{2}(1-\sqrt{5})$, corresponding to the overlap between $\vec{l}_1$ and $\vec{l}_3$, $\vec{l}_2$ and $\vec{l}_4$ etc. Also the input state subject to all the transformations has two components equal to $\pm\frac{1}{\sqrt[4]{5}}$ and one equal to $\pm\sqrt{1-\frac{1}{\sqrt{5}}}$. Keeping in mind that the same physical state corresponds to two opposite vectors (i. e. squared spin eigenstates are given by $\pm$-directions) we see that above transformations indeed realize the pentagram inequality.

The calibration of wave plates was performed modulo 90°. Rotation of a waveplate by ±90° corresponds to swapping its fast and slow axes. This transformation only results in a π phase-shift that the beam with the wave plate acquires relative to the other beam. This phase-shift is compensated automatically in our experimental procedure where we adjust the overall phase differences in the Mach-Zehnder interferometers.